\documentclass[prb,twocolumn,showpacs,preprintnumbers,amsmath,amssymb]{revtex4}

\usepackage{graphicx}
\usepackage{dcolumn}
\usepackage{bm}
\usepackage{subfigure}
\newcommand{\etal}{{\it et al.}}

\begin{document}


\title{Orbital currents, anapoles, and magnetic quadrupoles in CuO}
\author{S. Di Matteo$^{1}$ and M. R. Norman$^{2}$}
\affiliation{
$^1$ Groupe th\'eorie, D\'epartement Mat\'eriaux et Nanosciences, Institut de Physique de Rennes UMR UR1-CNRS 6251, Universit\'e de Rennes 1, F-35042 Rennes Cedex, France \\
$^2$ Materials Science Division, Argonne National Laboratory, Argonne, IL  60439, USA}

\date{\today}

\begin{abstract}

We show that orbital currents in a CuO$_2$ plane, if present, should be described by two independent parity and time-reversal odd order parameters, a toroidal dipole (anapole) and a magnetic quadrupole.  Based on this, we derive the resonant X-ray diffraction cross-section for monoclinic CuO at the antiferromagnetic wavevector and show that the two order parameters can be disentangled.  From our analysis, we examine a recent claim of detecting anapoles in CuO.

\end{abstract}

\maketitle


\section{Introduction}

A theory of the pseudogap phase in high temperature superconducting cuprates, proposed by Varma\cite{varma1,varma2}, introduced intra-unit cell orbital currents within CuO$_2$ planes as a key feature characterizing the physical properties of this phase. Varma's theory is based on a mean-field solution of a three band Hubbard model, and predicted two possible orbital current patterns, with their resulting orbital moments pointing out of the plane.  Photoemission dichroism \cite{adam} and neutron scattering experiments \cite{neutron1,neutron2,neutron3} are consistent with one of these  patterns, depicted in Fig.~\ref{twoloopA}.  This current pattern can be characterized by an in-plane vector order parameter, the toroidal moment (or anapole) \cite{matteo-varma}.  This vector, odd under inversion and time-reversal, has the same symmetry properties as the magnetoelectric susceptibility, which it is usually \cite{spaldin} identified with. For the CuO$_2$ case, the anapole has two possible in-plane orientations \cite{shvarma}, forming an E$_u$ representation.  But the neutron experiments \cite{neutron1} in addition show that the observed magnetic moments have an in-plane component, which is not consistent with solely in-plane currents if they have an orbital origin.  Very recently, a resonant X-ray diffraction (RXD)
measurement performed at the Cu L$_3$ edge also claimed the detection of orbital currents in the commensurate antiferromagnetic phase of monoclinic CuO, where the inferred anapole vector does not lie in the CuO$_2$ planes \cite{scagnoli}.

The aims of the present article are the following: (a) for a proper characterization of the orbital currents, a second order parameter, the magnetic quadrupole, is needed; (b) the direction of the inferred toroidal moment in CuO is actually 90$^\circ$ rotated from that illustrated in Ref.~\onlinecite{scagnoli}; (c) a toroidal origin for the observed RXD signal at the Cu L$_3$ edge in CuO is unlikely since the E1-M1 matrix elements are extremely small - rather, such effects would be more visible at the Cu K edge through E1-E2 interference. And although symmetry allowed, we believe anapoles
are more likely in the higher temperature multiferroic \cite{kimura} phase of CuO, where the moments form a helix \cite{brown}. Regardless, we emphasize the presence of a second, independent, order parameter, the magnetic quadrupole, that is time-reversal and parity odd, and has in principle the same order of magnitude as the toroidal dipole, and therefore should play a role in any description of orbital currents. Though the presence of the magnetic quadrupole does not change the theory of Ref.~\onlinecite{shvarma} qualitatively, some details have to be reconsidered. We conclude by highlighting some general ambiguities in the detection of the orbital currents by means of resonant X-ray diffraction.

\section{Toroidal dipole and magnetic quadrupole}

The toroidal dipole and magnetic quadrupole are intimately connected, being, respectively, the antisymmetric and the traceless symmetric parts of the bilinear form $r_{\alpha}m_{\beta}$, where $\vec{r}$ is the position vector of the magnetic moment $\vec{m}=\mu_B (\vec{L}+2\vec{S})$, with $\mu_B=\frac{e}{2m}$ the Bohr magneton in SI units and $\alpha,\beta = x,y,z$.
They can be defined from the expansion of the magnetic energy $W_m\equiv \int d\vec{r} \vec{J}(\vec{r})\cdot \vec{A}(\vec{r})$ up to second order in the field (here $\vec{J}(\vec{r})$ is the current density and $\vec{A}(\vec{r})$ the vector potential at position $\vec{r}$). This can be written as \cite{spaldin,dubovik,toprev}
$W_m=-\vec{m}\cdot\vec{B}+\frac{1}{2}\sum_{ij}M_{ij}\frac{\partial B_i}{\partial x_j}(0)+\vec{T}\cdot\vec{\nabla}\times\vec{B}$. Besides the magnetic moment $\vec{m}$, we defined the magnetic quadrupole $M_{ij}= \frac{1}{6} \int dV\left[3(x_im_j+m_ix_j) -2(\vec{x}\cdot\vec{m})\delta_{ij}\right]$ and the magnetic (polar) toroidal moment: $\vec{T}= \int dV  (\vec{x}\times \vec{m})$ \cite{foot1}.

From this, it is possible to identify the toroidal dipole $\vec{T}$ as the second order term (equivalent order as the magnetic quadrupole) that couples to the curl of the magnetic field (i.e., to currents and displacement currents). It is worthwhile to underline this result in the light of recent literature \cite{spaldin,toprev}: the presence of toroidal dipoles $\vec{T}$ does not necessarily imply a magnetoelectric coupling of the kind $\vec{T}\cdot(\vec{E}\times\vec{B})$. A symmetry analysis alone does not allow one to decide whether the $\vec{T}$-dipole is rather coupled to $\vec{\nabla}\times\vec{B}$, which is clearly not magnetoelectric.
This means that in a magnetoelectric material with an antisymmetric magnetoelectric tensor $\alpha_{ij} = - \alpha_{ji}$, one is not allowed to identify the 3 components of this latter with the 3 components of $\vec{T}$.
In fact, though $(\vec{\nabla}\times\vec{B})$ couples with $\vec{T}$, we are not guaranteed that $(\vec{E}\times\vec{B})$ couples with the same vector, i.e., with a current loop around a torus. 
Therefore, a material characterized by both a parity odd and a time-reversal odd symmetry does not necessarily couple to the vector $(\vec{E}\times\vec{B})$ in its free-energy expansion. A coupling with $(\vec{\nabla}\times\vec{B})$ respects these symmetries as well, without showing magnetoelectricity. As such, orbital currents, that can be described through $\vec{T}$ and $M_{ij}$, are not necessarily associated with magnetoelectricity.

\begin{figure}[b]
\includegraphics[width=3.0in]{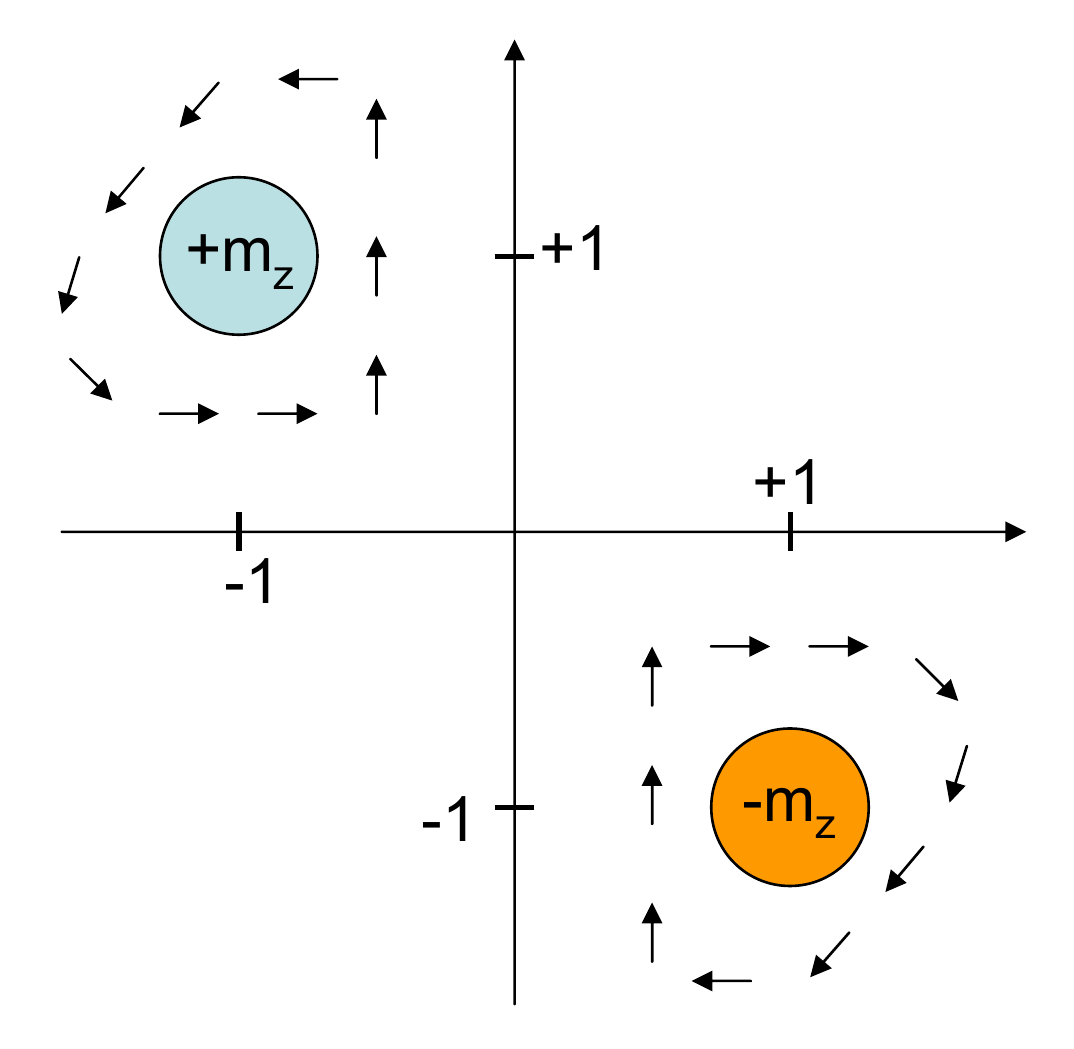}
\caption{(Color online) Two loop current pattern for CuO$_2$ planes \cite{varma1,varma2}.  The magnetic moment $\vec{m}$ at $\vec{r}$=(-1,1,0) is in the +$z$ direction and the magnetic moment at -$\vec{r}$=(1,-1,0) is in the -$z$ direction. This configuration, characterized by the bilinear $r_{\tilde{x}}m_z$ where $\tilde{x}$ is along (-1,1,0), is an equal admixture of a toroidal dipole and a magnetic quadrupole.}
\label{twoloopA}
\end{figure}

\begin{figure}[b]
\begin{center}
\includegraphics[width=3.0in]{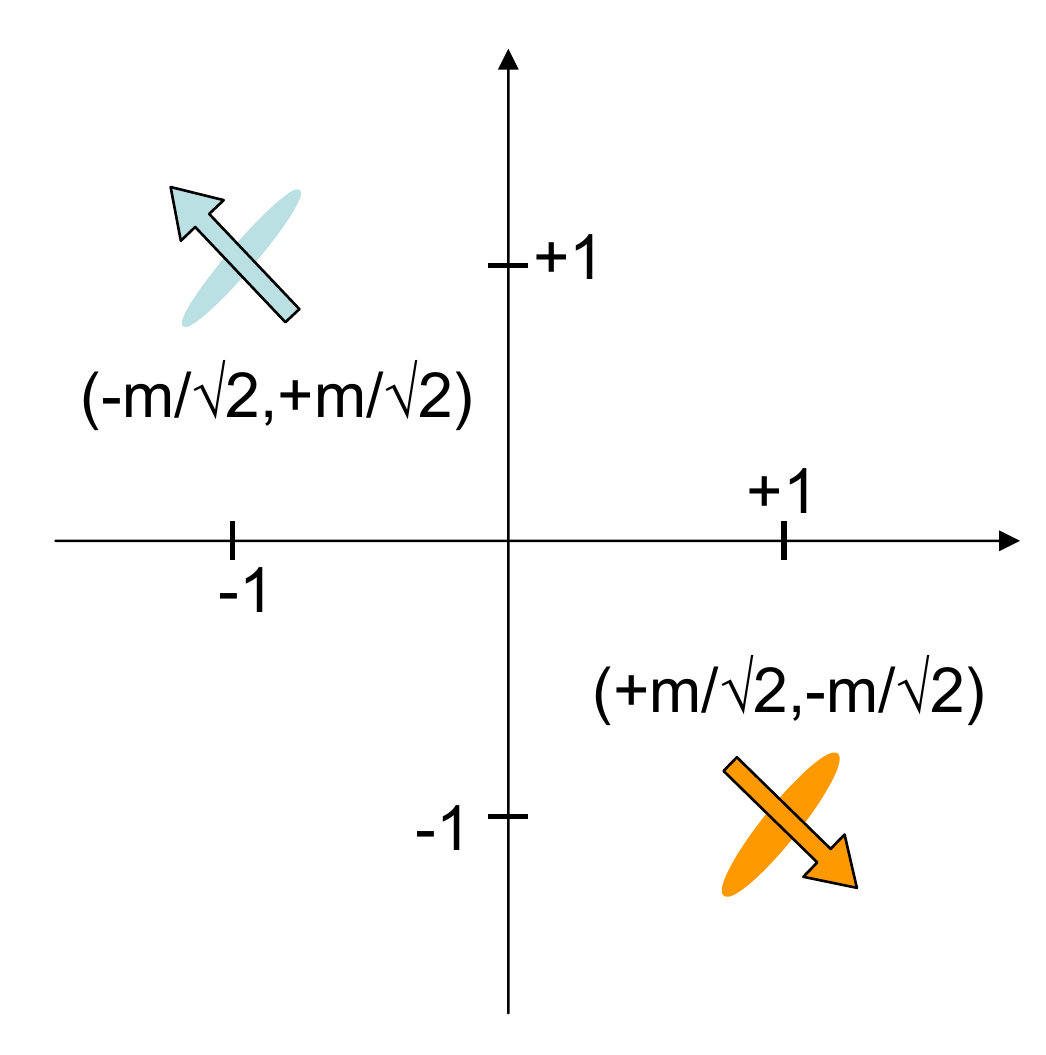}
\end{center}
\caption{(Color online) Two loop current pattern for CuO$_2$ planes: the pure magnetic quadrupole case. The magnetic moment  $\vec{m}$ at $\vec{r}$=(-1,1,0) is in the $xy$-plane with components $(-m/\sqrt{2},m/\sqrt{2},0)$ and the magnetic moment at -$\vec{r}$=(1,-1,0) is in the $xy$-plane with components $(m/\sqrt{2},-m/\sqrt{2},0)$. This configuration, characterized by the bilinear $r_{\tilde{x}}m_{\tilde{x}}$ where $\tilde{x}$ is along (-1,1,0), is a pure magnetic quadrupole.}
\label{twoloopB}
\end{figure}

In general, whatever configuration of two oppositely oriented orbital currents occurs, it is always characterized by a magnetic quadrupole that can be associated with their flow, as pictorially shown in Figs.~\ref{twoloopA} and \ref{twoloopB}.

Consider a reference coordinate system $xyz$ and a magnetic distribution whose total net magnetization is zero, like the ones depicted in Figs.~\ref{twoloopA} and \ref{twoloopB} (two magnetic moments $\vec{m}^{\alpha}$ at positions $\vec{r}^{\alpha}$, $\alpha=1,2$). It is possible to define the magnetic quadrupole and the magnetic toroidal dipole for these configurations as the first magnetic multipoles that are different from zero

\begin{align}
M_{ij} \equiv \sum_{\alpha=1,2} \left[  m^{\alpha}_{i} r^{\alpha}_{j} + r^{\alpha}_{i} m^{\alpha}_{j} - \frac{2}{3} \delta_{ij} \sum_{l} m^{\alpha}_{l} r^{\alpha}_{l} \right]
\label{magquaddef}
\end{align}

\begin{align}
T_{ij} \equiv \sum_{\alpha=1,2} [ m^{\alpha}_{i} r^{\alpha}_{j} - r^{\alpha}_{i} m^{\alpha}_{j} ]
\label{tordipdef}
\end{align}

The magnetic quadrupole is the traceless symmetric part of the cartesian tensor $m_{i} r_{j}$ and the toroidal dipole is the antisymmetric part (we can then call $T_z \equiv T_{xy}$, $T_y \equiv T_{zx}$ and $T_x \equiv T_{yz}$).

We can evaluate $M_{ij}$ and $T_{ij}$ for the orbital currents of Fig.~\ref{twoloopA}. The two loop currents are equivalent, by Ampere's theorem, to two magnetic moments $\vec{m}$ directed along $\pm z$ at positions $(-1,1,0)$ and $(1,-1,0)$, respectively. We get therefore $M_{xz}=-M_{yz}=T_{x}=T_{y}$. It should be noticed that by rotating the axes of Fig.~\ref{twoloopA} by 45$^\circ$, only one component of $M$ and one of $T$ differ from zero. This result is independent of the choice of the origin, as the toroidal dipole and the magnetic quadrupole are the first non-zero multipoles of this magnetic distribution. 

One might ask whether, by continuously deforming the relative positions of the two current loops, it is possible to find a configuration where the magnetic quadrupole is zero and only the toroidal dipole is present or vice versa.  The configuration of Fig.~\ref{twoloopA} is the only one having an equal value of toroidal dipole and magnetic quadrupole components, and any rotation of the magnetic dipoles (with the constraint that the total magnetic moment of the configuration is zero) leads to an increase of $M_{ij}$ and a decrease of $T$.  In the extreme case represented by the configuration of Fig.~\ref{twoloopB}, the toroidal dipole is zero.
Notice that there is no way to get a pure anapole from a two loop current. Indeed, it is not possible to get a pure anapole from any n-loop current, unless n goes to infinity and one gets a pure circulation of magnetic moments (equivalent to a current flowing around a torus). These considerations are important in order to build the correct order parameter for the two-loop current pattern of CuO$_2$ planes, by allowing out-of-plane components of the currents. In fact, the relative weight of $M_{ij}$ and $T_{ij}$ is a measure of the relative orientation of the current loops, as shown in two extreme cases in Figs.~\ref{twoloopA} and \ref{twoloopB}. One way to measure both order parameters is by means of RXD.

In the next section, we apply these considerations to the case of monoclinic CuO.

\section{Critical analysis of RXD in CuO}

Scagnoli \etal~measured the $\vec{Q}$=(1/2,0,-1/2) magnetic Bragg reflection in the low temperature monoclinic (Cc) phase of CuO at the Cu L$_3$ edge (i.e., in the commensurate antiferromagnetic phase). They found a dichroic signal with a $\sin(2\psi)$ dependence, where $\psi$ is the azimuthal angle about $\vec{Q}$. They interpreted their signal as evidence of E1-M1 (electric dipole-magnetic dipole) interference, which they related to the presence of a toroidal moment. We shall analyze this claim by repeating the calculations using a somewhat more transparent formalism based on cartesian tensors.

\begin{figure}[b]
\includegraphics[width=3.4in]{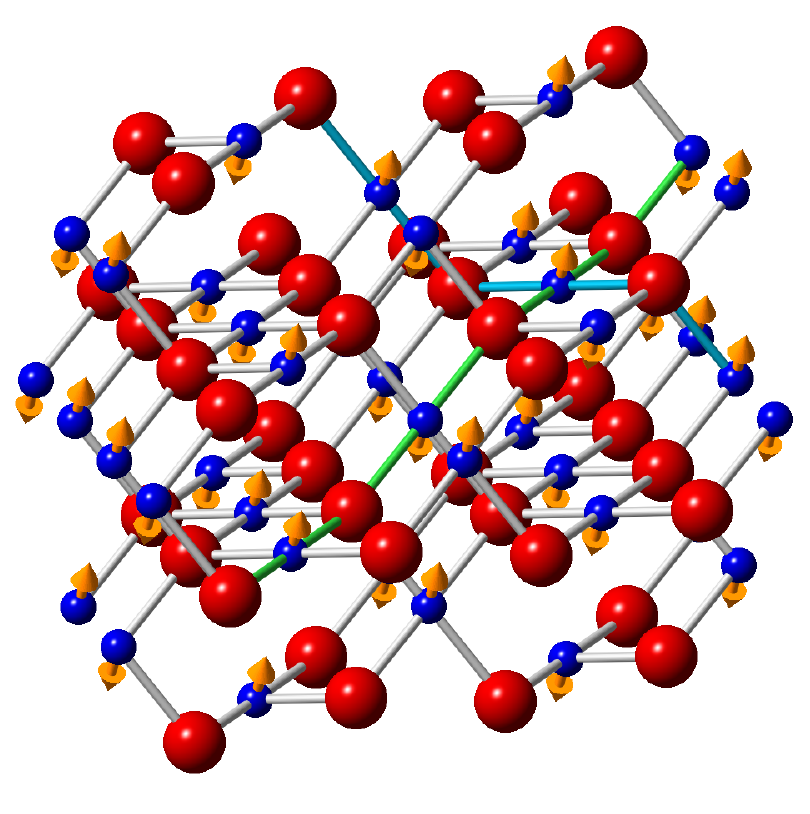}
\caption{(Color online) Commensurate antiferromagnetic structure of monoclinic CuO.  Small (blue) atoms are copper, large (red) oxygen, with arrows denoting the magnetic moments pointing along $\pm b$.  The antiferromagnetic chains run along (1,0,-1) (lower left to upper right; green), with intersecting (1,0,1) ferromagnetic chains (upper left to lower right; blue).  Ribbons of CuO$_2$ planes run along the (1,1,0) and (-1,1,0) directions.}
\label{conf1}
\end{figure}

\begin{figure}[b]
\includegraphics[width=3.4in]{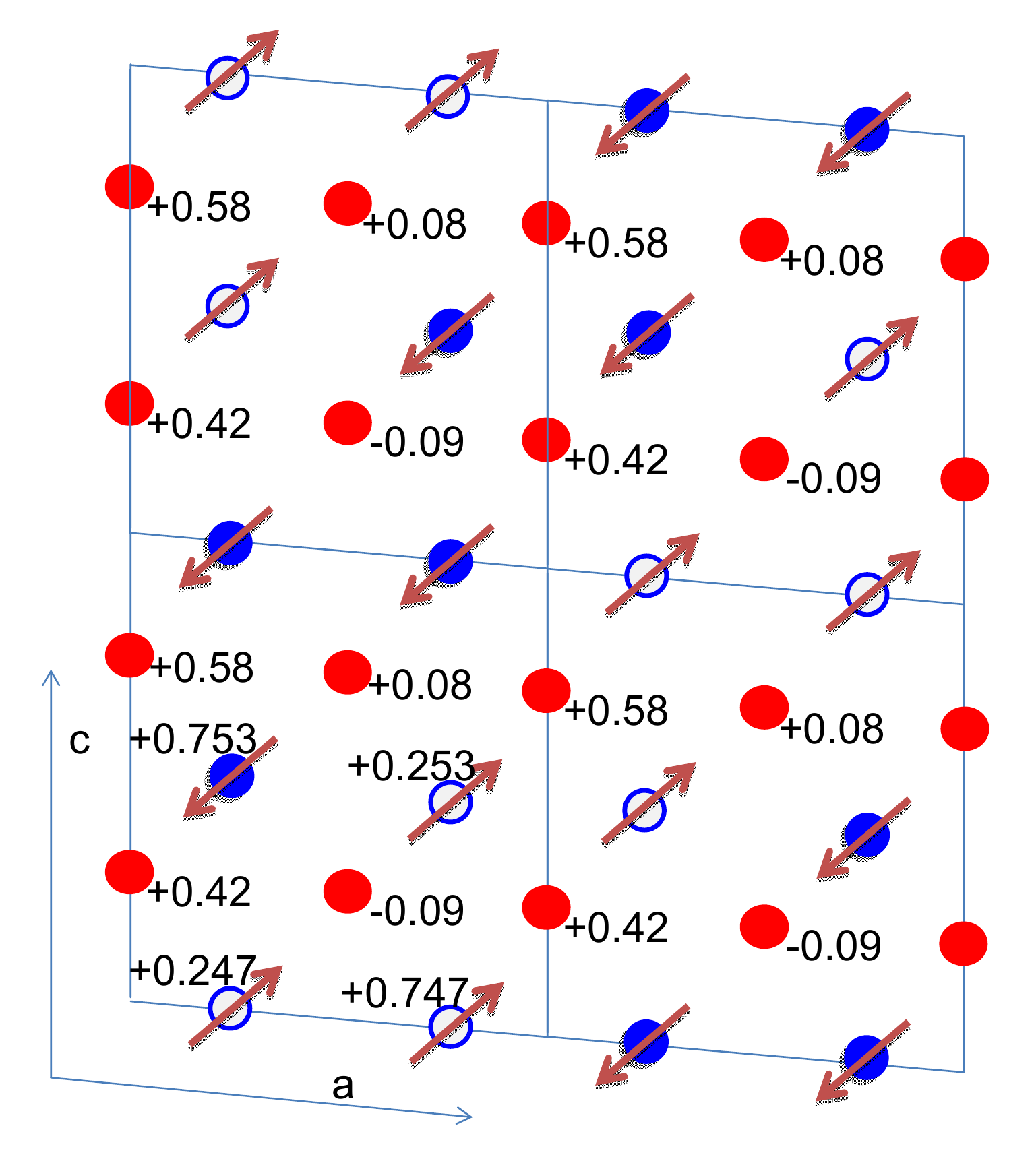}
\caption{(Color online) Commensurate antiferromagnetic CuO unit cell, equal to four times the crystallographic cell, projected in the $ac$-plane. The fractional $b$ coordinate of each ion is explicitly written. Open (blue) circles represent Cu ions having a magnetic moment pointing along $b$, closed (blue) circles along -$b$, with oxygen ions in the alternate rows (closed red circles). The anapoles (arrows) lie in the $ac$ plane, orthogonal to $\vec{Q}$.}
\label{conf2}
\end{figure}  

We begin with a description of the CuO commensurate antiferromagnetic structure, which is shown in Fig.~3.  This structure consists of (1,0,-1) antiferromagnetic chains which are intersected by (1,0,1) ferromagnetic chains.  Ribbons of CuO$_2$ planes run along the (1,1,0) and (-1,1,0) directions.  We describe the magnetic unit cell using 16 copper ions with doubled $a$ and $c$ lattice constants; therefore the $(1/2,0,-1/2)$ reflection becomes $(1,0,-1)$ in our notation. This unit cell corresponds to the four blocks shown in Fig.~4.  Copper ions are at the (equivalent) positions: ${\rm Cu}_1 = (1/8,y,0)$ - (${\hat E})$; ${\rm Cu}_2 = (3/8,y+1/2,0)$ - (${\hat E})$; ${\rm Cu}_3 = (5/8,y,0)$ - (${\hat T})$; ${\rm Cu}_4 = (7/8,y+1/2,0)$ - (${\hat T})$; ${\rm Cu}_5 = (1/8,\overline{y},1/4)$ - (${\hat T}{\hat m}_b)$; ${\rm Cu}_6 = (3/8,\overline{y}+1/2,1/4)$ - (${\hat m}_b)$; ${\rm Cu}_7 = (5/8,\overline{y},1/4)$ - (${\hat m}_b)$; ${\rm Cu}_8 = (7/8,\overline{y}+1/2,1/4)$ - (${\hat T}{\hat m}_b)$. In parenthesis, we included the symmetry with respect to the first ion ${\rm Cu}_1$: ${\hat E}$ is the identity, ${\hat T}$ is time-reversal, ${\hat m}_b$ is the glide plane perpendicular to the $b$-axis, and ${\hat T}{\hat m}_b$ is the product of the last two operations.
The 8 other ions are obtained by a translation of $(0,0,1/2)$, multiplied by  ${\hat T}$, as is clear from Fig.~4.

The structure factor is defined as $F(hkl) = \sum_{j=1}^{16} f_j(\omega) e^{2\pi i (hx_j+ky_j+lz_j)}$, where $x_j$, $y_j$ and $z_j$ are the three fractional coordinates for each of the 16 ions in the unit cell defined above, and $f_j(\omega)$ is the dynamical atomic scattering amplitude ($\omega$ is the photon energy) defined in Eq.~\ref{tmo1}. Here $(hkl)=(1,0,-1)$.  As all 16 Cu ions are related by some symmetry elements, the structure factor can be reduced to the dependence of only one atomic scattering factor \cite{carrathole,prlfink}, say that of $f_1$ (Cu$_1$). We get

\begin{eqnarray}
F & = & 2 e^{i\pi/4}(1+i)(1-{\hat T})(1+{\hat m}_b)f_1 
\label{strucfac}
\end{eqnarray}

The final result can be read as follows: the total amplitude (which must be squared to get the intensity) is provided by the time-reversal odd (magnetic, because of the term $(1-{\hat T})$) part of the atomic scattering factor of Cu$_1$ that is also even under the glide plane ${\hat m}_b$ (were it odd, the last factor $(1+{\hat m}_b)$ would have given zero). This result is important because it rules out a helical magnetic pattern arising from any residual component of the higher temperature multiferroic phase \cite{wu} as a cause of the signal.  In fact, only the component of the magnetic moment along the $b$ axis is detectable at the (1,0,-1) reflection, as the $a$ and $c$ components are odd under $\hat{m}_b$.

The atomic scattering factor $f_1$ appearing in Eq.~\ref{strucfac}, neglecting for the moment E2 (electric quadrupole) processes and considering as in Ref.~\onlinecite{scagnoli} just E1 and M1 terms, can be written as $f_1 \equiv f_1^{\rm{E1-E1}} + f_1^{\rm{E1-M1}}+ f_1^{\rm{M1-E1}} +f_1^{M1-M1}$. The order of magnitude of these terms has been calculated by means of the FDMNES program \cite{fdmnes} and M1 matrix elements have been found to be less than four orders of magnitude smaller than the E1 matrix elements. Therefore we can definitively neglect the $f_1^{M1-M1}$ amplitudes and the measured signal is therefore $I\propto |f_1^{\rm{E1-E1}} + f_1^{\rm{E1-M1}}+ f_1^{\rm{M1-E1}}|^2$. 

We define the above quantities as

\begin{eqnarray}
f_1^{\rm{E1-E1}} = \sum_n \Delta_n \langle \Psi_g | \vec{\epsilon}_o^* \cdot \vec{r} | \Psi_n \rangle \langle \Psi_n | \vec{\epsilon}_i \cdot \vec{r} |\Psi_g \rangle
\label{tmo1}
\end{eqnarray}

\noindent where the resonant denominator $\Delta_n = (\hbar \omega - (E_n - E_g))^{-1}$ provides the $\omega$ dependence (here we neglected damping, not important for symmetry considerations) and $\epsilon_{o,i}$ are the outgoing and incoming polarizations, respectively.
It is common practice to expand the two scalar products and then factorize the terms depending on the electromagnetic wave (the polarization in the E1-E1 case), and those depending on matter. This is usually done in spherical tensors \cite{matteo-varma,natoli05}, but it can be done in cartesian tensors as well \cite{jolyprb}.  We get

\begin{eqnarray}
f_1^{\rm{E1-E1}} = \sum_{\alpha,\beta} T_{\alpha\beta} F_{\alpha\beta}
\label{tmo1bis}
\end{eqnarray} 

\noindent where $T_{\alpha\beta}=(\vec{\epsilon}_o^*)_{\alpha}(\vec{\epsilon}_i)_{\beta}$ represents the electromagnetic wave and $F_{\alpha\beta}=\sum_n \Delta_n \langle \Psi_g | \vec{r}_{\alpha} | \Psi_n \rangle \langle \Psi_n |\vec{r}_{\beta}|\Psi_g \rangle$ represents the properties of the sample, with $\alpha,\beta=x,y,z$.  
With the same notation, we can write

\begin{align} \label{tmo2bis}
& f_1^{\rm{E1-M1}}+f_1^{\rm{M1-E1}} = \sum_n \Delta_n \\
& \left[ \langle \Psi_g | \vec{\epsilon}_o^* \cdot \vec{r} | \Psi_n \rangle \langle \Psi_n | \big(\vec{k}_i \times \vec{\epsilon}_i\big) \cdot \big(\vec{L}+2\vec{S}\big) |\Psi_g \rangle \right. \nonumber \\
& \left. + \langle \Psi_g | \big(\vec{k}_o \times \vec{\epsilon}_o^*\big) \cdot \big(\vec{L}+2\vec{S}\big) |\Psi_n \rangle \langle \Psi_n | \vec{\epsilon}_i \cdot \vec{r} |\Psi_g \rangle \right] =  \nonumber \\
& = \sum_{\alpha,\beta} \left[ (\vec{\epsilon}_o^*)_{\alpha}\big(\vec{k}_i \times \vec{\epsilon}_i\big)_{\beta} S_{\alpha\beta} + \big(\vec{k}_o \times \vec{\epsilon}_o^*\big)_{\alpha} (\vec{\epsilon}_i)_{\beta} {\tilde S}_{\alpha\beta} \right] \nonumber \\
& = \sum_{\alpha,\beta} \left[ (\vec{\epsilon}_o^*)_{\alpha}\big(\vec{k}_i \times \vec{\epsilon}_i\big)_{\beta} \big(\Re S_{\alpha\beta}+i \Im S_{\alpha\beta} \big) \right. \nonumber \\
& \left. + \big(\vec{k}_o \times \vec{\epsilon}_o^*\big)_{\alpha} (\vec{\epsilon}_i)_{\beta} \big(\Re S_{\beta\alpha} - i\Im S_{\beta\alpha}\big) \right] \nonumber \\
& = \sum_{\alpha,\beta} \left[ \Re S_{\alpha\beta} \big( (\vec{\epsilon}_o^*)_{\alpha}\big(\vec{k}_i \times \vec{\epsilon}_i\big)_{\beta} +(\vec{\epsilon}_i)_{\alpha}\big(\vec{k}_o \times \vec{\epsilon}_o^*\big)_{\beta} \big) \right. \nonumber \\
& \left. + i\Im S_{\alpha\beta} \big( (\vec{\epsilon}_o^*)_{\alpha}\big(\vec{k}_i \times \vec{\epsilon}_i\big)_{\beta} -(\vec{\epsilon}_i)_{\alpha}\big(\vec{k}_o \times \vec{\epsilon}_o^*\big)_{\beta} \big) \right] \nonumber
\end{align}

\noindent where we defined the two hermitian conjugate quantities $S_{\alpha\beta}=\sum_n \Delta_n \langle \Psi_g | (\vec{r})_{\alpha} | \Psi_n \rangle \langle \Psi_n | \big(\vec{L}+2\vec{S}\big)_{\beta} |\Psi_g \rangle$ and ${\tilde S}_{\alpha\beta}=\sum_n \Delta_n \langle \Psi_g | \big(\vec{L}+2\vec{S}\big)_{\alpha} |\Psi_n \rangle \langle \Psi_n | (\vec{r})_{\beta} |\Psi_g \rangle$. Notice that for the E1-E1 term, the antisymmetric and complex conjugate coincide for $F_{\alpha\beta}$, but this is not the same for $S_{\alpha\beta}$ and ${\tilde S}_{\alpha\beta}$: $\Re [S_{\alpha\beta}] = \Re [{\tilde S}_{\beta\alpha}]$ and $\Im [S_{\alpha\beta}] = -\Im [{\tilde S}_{\beta\alpha}]$. This relation was used in the next to last step of Eq.~\ref{tmo2bis}.

There are two independent, real 3x3 matrices appearing in Eq.~\ref{tmo2bis}, $\Re S_{\alpha,\beta}$, time-reversal odd, and $i\Im S_{\alpha,\beta}$, time-reversal even, as can be seen from their definitions

\begin{align}
\Re S_{\alpha\beta} & \equiv \frac{1}{2} \sum_n \Delta_n \left[  \langle \Psi_g | (\vec{r})_{\alpha} | \Psi_n \rangle \langle \Psi_n | \big(\vec{L}+2\vec{S}\big)_{\beta} |\Psi_g \rangle \right. \nonumber \\
& \left. + \langle \Psi_g | \big(\vec{L}+2\vec{S}\big)_{\beta} | \Psi_n \rangle \langle \Psi_n | (\vec{r})_{\alpha}|\Psi_g \rangle \right]
\label{todd}
\end{align}

\begin{align}
i\Im S_{\alpha\beta} & \equiv \frac{1}{2} \sum_n \Delta_n \left[  \langle \Psi_g | (\vec{r})_{\alpha} | \Psi_n \rangle \langle \Psi_n | \big(\vec{L}+2\vec{S}\big)_{\beta} |\Psi_g \rangle \right. \nonumber \\
& \left. - \langle \Psi_g | \big(\vec{L}+2\vec{S}\big)_{\beta} | \Psi_n \rangle \langle \Psi_n | (\vec{r})_{\alpha}|\Psi_g \rangle \right]
\label{teven}
\end{align}

\noindent We remind that the time-reversal operation performs a complex conjugation {\it and} reverses the sign of $\vec{L}$ and $\vec{S}$. Notice the difference with the E1-E1 3x3 tensor, characterized by just 9 independent components, such that the symmetric (real) elements are time-reversal even and the antisymmetric (imaginary) elements are time-reversal odd. Instead, in the E1-M1 case, there are 18 terms, 9 of which are time-reversal even (imaginary, independently of spatial symmetry) and 9 are time-reversal odd (real, again independently of spatial symmetry). With spherical tensors, one can identify each of the 9 components in terms of irreducible tensors as ${\cal L}=0$ (1 component), ${\cal L}=1$ (3 components) and ${\cal L}=2$ (5 components). The correspondence of these irreducible tensors with cartesian tensors is very simple:  ${\cal L}=0$ corresponds to the trace of the matrix,  ${\cal L}=1$ is the antisymmetric part, ${\cal L}=2$ is the traceless symmetric part. This is valid for all the E1-E1 amplitudes and each of the time-reversal odd/even matrices of E1-M1 events.

In the light of the experiment on CuO, we focus on time-reversal odd terms, $\Re S_{\alpha\beta}$, and use the following spherical linear combinations:

\begin{equation}
\big[ \Re S \big]^{(0)} \equiv \sum_{\alpha} \Re S_{\alpha\alpha}
\label{scalar}
\end{equation}
\begin{equation}
\big[ \Re S \big]^{(1)} \equiv \frac{1}{2} \left[  \Re S_{\alpha\beta} - \Re S_{\beta\alpha} \right]
\label{antisym}
\end{equation}
\begin{equation}
\big[ \Re S \big]^{(2)} \equiv \frac{1}{2} \left[ \Re S_{\alpha\beta} + \Re S_{\beta\alpha} -\frac{2}{3} \delta_{\alpha\beta} \Re S_{\alpha\beta} \right]
\label{sym}
\end{equation}

\noindent The pseudoscalar part (Eq.~\ref{scalar}) is not effective in our case as it is glide-plane odd.
The antisymmetric term (Eq.~\ref{antisym}) is the anapole (polar toroidal dipole), whereas the symmetric, rank-two tensor (Eq.~\ref{sym}) is the magnetic quadrupole.

\begin{figure}[b]
\includegraphics[width=3.4in]{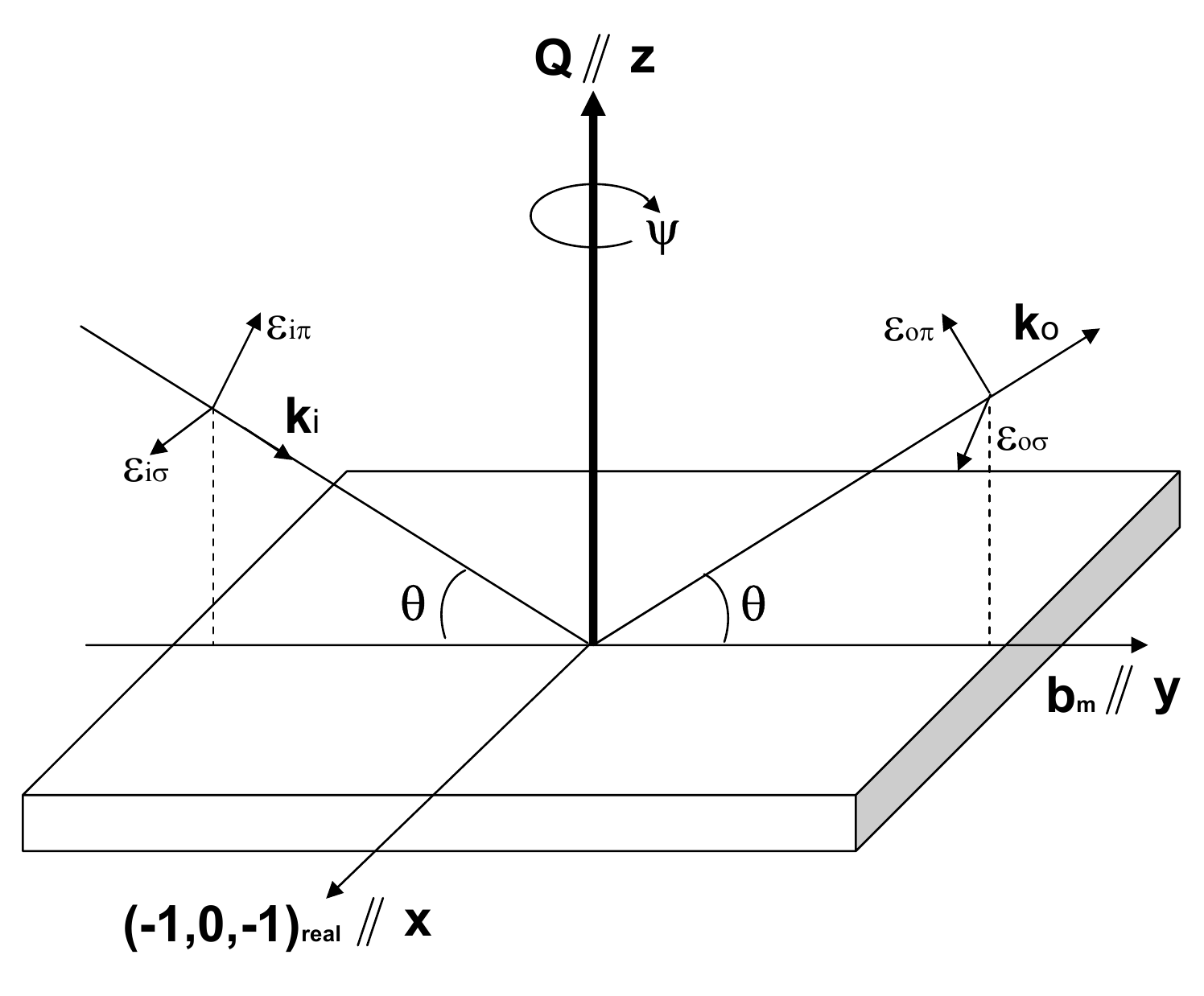}
\caption{Diffraction plane defined by the vectors $\vec{k}_i$ and $\vec{k}_o$. It is drawn for $\psi=0$, when the diffraction plane contains the $\vec{b}$ monoclinic axis.}
\label{diffplane}
\end{figure}  

For actual calculations, we choose a reference frame with the $z$ direction along $\vec{Q}=(1,0,-1)$, the $y$ direction along the $b$ axis, and the $x$ direction along the $(-1,0,-1)$ direction (real space), in order to have a right-handed frame. In our frame, the sample is fixed and the x-ray beam rotates clockwise, in order to match the experimental condition where the beam is fixed and the sample rotates counterclockwise. According to the choice of Scagnoli \etal, the azimuthal angle $\psi$ is zero when the diffraction plane contains the $b$ monoclinic axis (see Fig.~\ref{diffplane}). An intrinsic ambiguity of sign (both $\psi$ and $\psi+\pi$ satisfy this condition) does not affect the final results for the anapole, but can affect the sign of the magnetic quadrupole contribution (see Eq.~\ref{dichroism3} below).
Some definitions useful for the following, in our $xyz$ frame of Fig.~\ref{diffplane}, are:
\begin{eqnarray}
\vec{\epsilon}_{i\sigma} & = & \vec{\epsilon}_{o\sigma}=(-\cos\psi,\sin\psi,0) \nonumber \\
\vec{\epsilon}_{i\pi} & = & (\sin\theta\sin\psi, \sin\theta\cos\psi, \cos\theta) \nonumber \\
\vec{\epsilon}_{o\pi} & = & (-\sin\theta\sin\psi, -\sin\theta\cos\psi, \cos\theta) \nonumber \\
\vec{k}_{i} & = & k(\cos\theta\sin\psi, \cos\theta\cos\psi, -\sin\theta) \nonumber \\
\vec{k}_{o} & = & k(\cos\theta\sin\psi, \cos\theta\cos\psi, \sin\theta) \nonumber \\
\vec{Q} & \equiv & \vec{k}_{o} - \vec{k}_{i} = 2k(0,0, \sin\theta) \nonumber \\
\vec{k}_{i} + \vec{k}_{o} & = & 2k(\cos\theta\sin\psi, \cos\theta\cos\psi,0) \nonumber
\end{eqnarray}
\noindent where $k$ is the modulus of the wavevector and $\theta$ is the Bragg angle.

The E1-E1 atomic scattering factor is known to be\cite{luo}

\begin{eqnarray}
f_1^{\rm{E1-E1}}\propto i \left[\vec{\epsilon}_o^* \times \vec{\epsilon}_i\right] \cdot \vec{m}
\label{E1E1}
\end{eqnarray}

Instead, the E1-M1 atomic scattering amplitudes contributing to the structure factor (\ref{strucfac}) are the components of the anapole in the $xz$ plane and the $xy$ and $yz$ components of the magnetic quadrupole. The reason is that the anapole is a time-reversal odd, polar vector, and therefore its $y$ component is odd under $\hat{m}_b$, contrary to the $x$ and $z$ components. The magnetic quadrupole is such that only the $xz$ and $yz$ components are non zero after the application of $1+\hat{m}_b$ in Eq.~\ref{strucfac}. The contribution proportional to the anapole is the vector term from Eq.~\ref{antisym}, whose polarization and wavevector components are given by $\vec{P} \equiv (\vec{\epsilon}_o^*)\times \big(\vec{k}_i \times \vec{\epsilon}_i\big) - \big(\vec{k}_o \times \vec{\epsilon}_o^*\big) \times (\vec{\epsilon}_i)$. Therefore Eq.~\ref{tmo2bis}, limited to this term, becomes: $f_{1a}  \propto \vec{T} \cdot \vec{P}$, where $\vec{T}$ is the toroidal dipole (anapole) vector \cite{foot2}. 
For the magnetic quadrupole, similar rules apply. In particular, for the $M_{rs}$ component, we have $f_{1b}  \propto M_{rs} \big( (\vec{\epsilon}_o^*)_r (\vec{k}_i \times \vec{\epsilon}_i)_s + (\vec{\epsilon}_o^*)_s (\vec{k}_i \times \vec{\epsilon}_i)_r + (\vec{\epsilon}_i)_r (\vec{k}_o \times \vec{\epsilon}_o^*)_s + (\vec{\epsilon}_i)_s (\vec{k}_o \times \vec{\epsilon}_o^*)_r \big)$. 

From these definitions we can evaluate the scattering amplitudes in the $\sigma$, $\pi$ and circular $\pm$ polarization conditions. The E1-E1 magnetic structure factors are
\begin{eqnarray}
f_1^{\rm{E1-E1}}(\sigma \sigma) & = & 0 \nonumber \\
f_1^{\rm{E1-E1}}(\sigma \pi) & \propto & -im_y\cos\theta \cos\psi \nonumber \\
f_1^{\rm{E1-E1}}(\pi \sigma) & \propto & im_y\cos\theta \cos\psi \nonumber \\
f_1^{\rm{E1-E1}}(\pi \pi) & \propto & im_y\sin(2\theta) \sin\psi
\label{fE1E1}
\end{eqnarray}

whereas those corresponding to $f_{1a}$ are:
\begin{eqnarray}
f_{1a}(\sigma \sigma) & \propto & 2T_x\cos\theta\sin\psi \nonumber \\
f_{1a}(\sigma \pi) & \propto & -T_x\sin(2\theta) \cos\psi \nonumber \\
f_{1a}(\pi \sigma) & \propto & T_x\sin(2\theta) \cos\psi \nonumber \\
f_{1a}(\pi \pi) & \propto & 2T_x\cos\theta \sin\psi
\label{fE1M1a}
\end{eqnarray}

and those corresponding to $f_{1b}$ are:
\begin{eqnarray}
f_{1b}(\sigma \sigma) & \propto & 2M_{yz}\cos\theta\sin\psi \nonumber \\
f_{1b}(\sigma \pi) & \propto & M_{xy}\cos^2\theta\sin(2\psi)  \nonumber \\
f_{1b}(\pi \sigma) & \propto & M_{xy}\cos^2\theta\sin(2\psi)  \nonumber \\
f_{1b}(\pi \pi) & \propto & -2M_{yz}\cos\theta \sin\psi
\label{fE1M1b}
\end{eqnarray}

Amplitudes for circular polarizations are defined as $f_C^{\pm}\equiv f_{\sigma\sigma}\pm i f_{\sigma\pi}$. Circular dichroism can be defined in two ways, according to whether one measures or not the outgoing polarizations. In the former case if (say) outgoing measured polarization is $\sigma$, one should evaluate $\Delta I_{\sigma} \equiv I_{+,\sigma} - I_{-,\sigma} $. If no outgoing polarization is measured, then $\Delta I \equiv I_{+,\sigma} + I_{+,\pi} - I_{-,\sigma} - I_{-,\pi} $ should be evaluated. The latter case is that of Ref.~\onlinecite{scagnoli} \cite{staub}. From the previous definitions we get:

\begin{eqnarray} \label{dichroism}
\Delta I & = & \left| F_{+,\sigma} \right| ^2 + \left| F_{+,\pi} \right| ^2 - \left| F_{-,\sigma} \right| ^2 - \left| F_{-,\pi} \right| ^2 \\
& \propto & \left| f_{\sigma,\sigma} + i f_{\pi,\sigma}\right| ^2 + \left| f_{\sigma,\pi} + i f_{\pi,\pi}\right| ^2 \nonumber \\
&  -  &\left| f_{\sigma,\sigma} - i f_{\pi,\sigma}\right| ^2 - \left| f_{\sigma,\pi} - i f_{\pi,\pi}\right| ^2 \nonumber \\
& \propto & - \Re f_{\sigma,\sigma} \Im f_{\pi,\sigma} - \Re f_{\sigma,\pi} \Im f_{\pi,\pi} \nonumber \\
& + & \Re f_{\pi,\sigma} \Im f_{\sigma,\sigma} + \Re f_{\pi,\pi} \Im f_{\sigma,\pi} \nonumber 
\end{eqnarray}

Analogously: 
\begin{eqnarray} \label{dichroism2}
\Delta I_{\sigma} = \left| F_{+,\sigma} \right| ^2 - \left| F_{-,\sigma} \right| ^2 
\propto - \Re f_{\sigma,\sigma} \Im f_{\pi,\sigma} + \Re f_{\pi,\sigma} \Im f_{\sigma,\sigma} \nonumber \\
\end{eqnarray}

Taking the imaginary quantities from Eq.~\ref{fE1E1}, and the real quantities from Eqs.~\ref{fE1M1a} and \ref{fE1M1b}, we get for the first dichroism:

\begin{eqnarray}
\Delta I & \propto & m_y T_x \cos^4\theta  \sin(2\psi) \nonumber \\
 &  +  & m_y M_{xy} \cos^3\theta \sin\theta  \sin(2\psi)\sin\psi
\label{dichroism3}
\end{eqnarray}

and for the second:
\begin{eqnarray}
\Delta I_{\sigma} \propto m_y (T_x + M_{yz}) \cos^2\theta  \sin(2\psi) 
\label{dichroism4}
\end{eqnarray}

Therefore from a coupled analysis of the azimuthal scan of both kinds of dichroism, it is possible to obtain the relative value of all three components $T_x$, $M_{xy}$ and $M_{yz}$: the $\sin\psi$ correction to the $\sin(2\psi)$ azimuthal scan is proportional to $M_{xy}$ (if the values of $m_y$ and of the radial transition matrix elements are known). More interesting is the information that we can get from the ratio of the $\sin(2\psi)$ modulation of $\Delta I$ and $\Delta I_{\sigma}$: $4\Delta I_{\sigma}/\Delta I = \cos^{-2}\theta (1+M_{yz}/T_x)$, which is an absolute measurement of the magnetic quadrupole to toroidal dipole ratio, independent of $m_y$ and of any radial transition matrix element. This would be important in the light of the findings of Section II, concerning the ratio of the anapole to magnetic quadrupole as a measure of the out-of-plane component of the orbital currents.

We also performed numerical calculations by means of the FDMNES program at the Cu pre K-edge at the same reflection. The aim is to find out whether clearer evidence of orbital currents can be obtained through E1-E2 matrix elements, e.g., transitions from $1s$ to $4p$ levels and then from $3d$ back to $1s$, by taking advantage of the $pd$ hybridization in CuO. The calculated order of magnitude of the signal in the $\sigma\sigma$ channel, though small $\sim 4 \cdot 10^{-4} r_0^2$, is well within the present sensitivity at synchrotron facilities. However, several multipole components are present, as detailed below, and in order to identify each term, one should take advantage of both azimuthal scans and measurements of several reflections of the same class as $(1,0,-1)$, in order to change the value of $\theta$ as well. 
The measured components at the Cu pre K-edge are the following: in the E1-E2 channel we get $T_x$, $T_z$, $M_{xy}$, $M_{yz}$, $T_{z^3}$, $T_{xz^2}$, $T_{xyz}$, $T_{x(x^2-3y^2)}$, where $T_{\alpha\beta\gamma}$ is the toroidal octupole.
Moreover, at the pre K-edge, E2-E2 terms in the $\sigma \sigma$ channel are present, of the same order of magnitude ($\sim 10^{-4} r_0^2$). They are: $m_y$, $O_{yz^2}$, $O_{z(x^2-y^2)}$ and $O_{y(3x^2-y^2)}$, where $O_{\alpha\beta\gamma}$ is the magnetic octupole.
Focusing on just the toroidal dipole and magnetic quadrupole, their azimuthal dependence is the following \cite{carra,lovesey}: 

\begin{eqnarray}
f_{K}(\sigma \sigma) & \propto &2 \cos\theta\sin\psi \left[T_x+\frac{\sqrt{5}}{3}M_{yz}\right]  \nonumber \\
f_{K}(\sigma \pi) & \propto &  \sin(2\theta)\cos\psi \left[T_x-\frac{2\sqrt{5}}{3}M_{yz}\right]   \nonumber \\
&  & -\frac{\sqrt{5}}{3}\cos^2\theta\sin(2\psi)M_{xy} \nonumber \\
f_{K}(\pi \sigma) & \propto &  -  \sin(2\theta)\cos\psi \left[T_x-\frac{2\sqrt{5}}{3}M_{yz}\right] \nonumber \\
&  & -\frac{\sqrt{5}}{3}\cos^2\theta\sin(2\psi)M_{xy}\nonumber \\
f_{K}(\pi \pi) & \propto &  2\cos\theta\sin\psi \left[T_x-\frac{\sqrt{5}}{3}M_{yz}\right]\nonumber \\
&  & \left(\cos^2\theta-3\sin^2\theta \right)
\label{kedge}
\end{eqnarray}

This could be compared to future experiments.
Interestingly, the full structure factor for a generic $(h,0,l)$ reflection reads: 

\begin{eqnarray}
F(h0l) & = & e^{i\pi/4}(1+(-)^h{\hat T})(1+(-)^l{\hat T})(1+(-i)^h{\hat T}) \nonumber \\
&  &  (1+(i)^{h+l}{\hat m}_b)f_1 
\label{strucfac2}
\end{eqnarray}

\noindent an expression that reduces to Eq.~(\ref{strucfac}) when $h=1$ and $l=-1$. This shows that magnetic reflections are characterized by both $h$ odd and $l$ odd. The absence of magnetic reflection as measured by neutron scattering for $h+l=4n+2$ showed that the magnetic moment has only the $y$ component (otherwise the glide plane would have implied a canted $x$ or $z$ component that would have given an intensity to this class of reflections). However, this extinction rule, implied by the $(1+(i)^{h+l}{\hat m}_b)$ term, might be violated for E1-E2 terms. A signal at, e.g., the $(1,0,1)$ reflection is in this case a direct indication of a magnetic quadrupole or toroidal dipole and should be looked for.

\section{Conclusions}

Given the smallness of the M1 matrix elements, we find that the dichroism signal observed in Ref.~\onlinecite{scagnoli} is unlikely to be due to E1-M1 interference.  And although anapoles are symmetry allowed for the commensurate antiferromagnetic phase of CuO, we find them physically unlikely, as the displacement of the Cu ions off of their C2/c locations in the Cc structure is along $b$,\cite{asbrink} which is the direction of the magnetic moments.  The fact that the resulting cross product is zero casts doubts on the existence of anapoles arising from a magnetoelectric effect. 

Recently, Joly has suggested that the $\sigma\sigma$ signal observed by Scagnoli \etal \cite{scagnoli} is most likely due to birefringence \cite{private}.  That is, because of the biaxial nature of the monoclinic structure, the light will rotate in the sample, causing a $\sigma\pi$ signal of purely magnetic dipole origin to show up in the $\sigma\sigma$ channel.  In fact, the split nature of the RXD peak observed in Ref.~\onlinecite{scagnoli} is almost certainly due to strong self-absorption, leading further credence to this scenario.  Whether such a scenario can account for the pronounced $\sin(2\psi)$ azimuthal dependence of the observed dichroism signal remains to be seen.

From the above considerations, it would be extremely interesting to look at RXD in the multiferroic phase, especially at the pre K-edge, where  birefringence effects should be much less important. In this phase, the magnetic moments rotate in a plane that contains $b$ and a vector near (1,0,3) \cite{brown}, with a ferroelectric polarization along $b$ \cite{kimura}.  Moreover, such experiments in the cuprates would be very illuminating in resolving the question of whether an orbital current description is appropriate for the observed neutron scattering signal in the pseudogap phase.

\begin{acknowledgments}

The authors thank Peter Abbamonte, Yves Joly and Urs Staub for various discussions. This work is supported by the US DOE, Office of Science, under Contract DE-AC02-06CH11357.

\end{acknowledgments}

\end{document}